\documentclass[twocolumn,showpacs,prl,superscriptaddress,preprintnumbers]{revtex4}
\usepackage{graphicx}
\def\Dirac#1{#1\hskip-5pt/}
\begin{document}
\title{Double deeply virtual Compton scattering off the nucleon}
\author{M. Guidal} 
\affiliation{Institut de Physique Nucl\'eaire, F-91406 Orsay, France}
\author{M. Vanderhaeghen}
\affiliation{Institut f\"ur Kernphysik, Johannes Gutenberg Universit\"at,
  D-55099 Mainz, Germany}
%
%
\begin{abstract}
We study the double deeply virtual Compton scattering (DDVCS) process off the
nucleon, through the scattering of a spacelike virtual photon with
large virtuality resulting in the production of a timelike virtual photon,
decaying into an $e^+e^-$ pair. This process is expressed in the Bjorken
regime in terms of generalized parton distributions (GPDs) and it is
shown that by varying the invariant mass of the lepton pair, one can 
directly extract the GPDs from the observables. 
We give predictions for the DDVCS cross
section and beam helicity asymmetry and discuss its
experimental feasibility.
\end{abstract}
\maketitle
\begin{figure}[h]
\vspace{-3.cm}
\includegraphics[width=8.5cm,height=12cm]{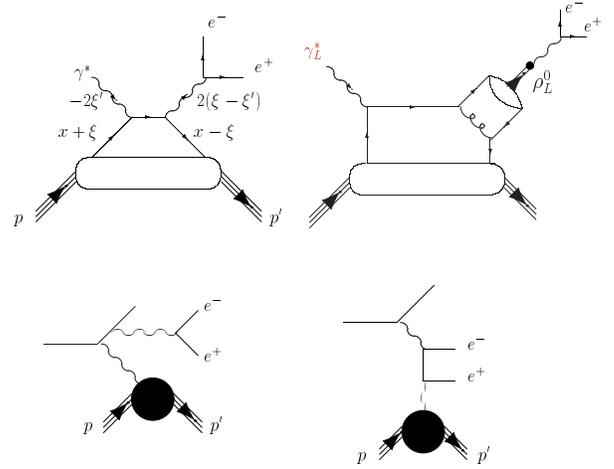}
\vspace{-3.25cm}
\caption[]{Diagrams for the $l p \to l p e^+ e^-$
process : DDVCS process (upper left), vector meson (VM) production process
(upper right), Bethe-Heitler (BH) processes (lower two diagrams). 
Crossed diagrams are not shown but also included.}
\label{fig:diagrams}
\end{figure}

\begin{figure}[h]
\vspace{-1.cm}
\includegraphics[width=6.65cm]{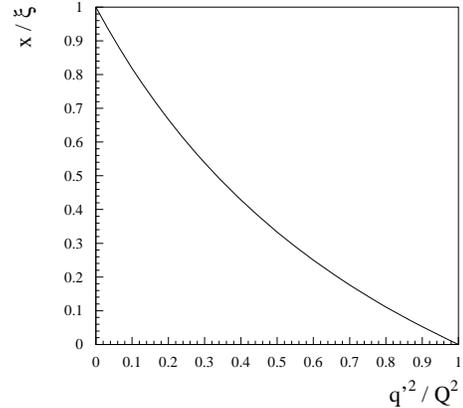}
\vspace{-0.5cm}
\caption[]{Range in the argument $x = 2 \xi^\prime - \xi$ (relative to $\xi$) 
of the GPD $(x, \xi, t)$ which one accesses by measuring 
the imaginary part of the 
DDVCS process at different values of the lepton invariant mass
$q'^2$ (relative to the initial virtuality $Q^2$).}
\label{fig:range}
\end{figure}

\begin{figure}[h]
\includegraphics[width=7.4cm]{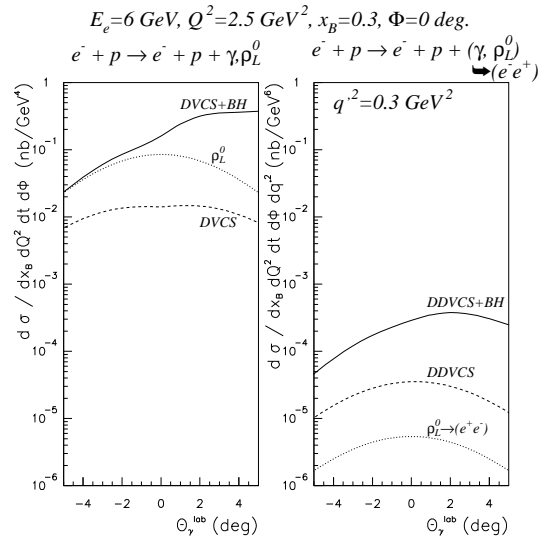}
\vspace{-0.4cm}
\caption[]{\small Comparison of the different cross sections (as
indicated on the curves) for the 
$e p \to e p (\gamma, \rho^0_L)$ (left panel) and 
$e p \to e p e^+ e^-$ (right panel) reactions 
in JLab kinematics.}
\label{fig:ddvcs6_qps}
\end{figure}

 
%
%
The understanding of hadron structure in terms of quark and
gluon degrees of freedom, remains an outstanding challenge. An
important source of information is provided 
by experiments involving electroweak probes. In this way, elastic 
form factors as well as quark and gluon distributions in
the nucleon have been mapped out in quite some detail. In recent
years, a whole new class of hard exclusive reactions have become 
accessible both theoretically and experimentally to study hadron structure. 
In particular, the deeply
virtual Compton scattering (DVCS) and hard electroproduction of meson 
processes are at present under investigation at different facilities 
(HERMES \cite{Air01}, JLab \cite{Step01}, HERA \cite{Adl01,Sau00}), 
or will be addressed by experiments in the near future.
In these processes, a highly virtual photon (with large virtuality $Q^2$)
scatters from the nucleon and a real photon (in the case of DVCS) or a
meson is produced. Due to the large scale $Q^2$ involved, these hard
exclusive processes are factorizable in a hard part, which can be
calculated from perturbative QCD, and a soft part, which contains the
information on nucleon structure and is parametrized in terms of
generalized parton distributions (GPDs) (see
Refs.~\cite{Ji98,Rad01,GPV01} for reviews and references therein).
\newline
\indent
The GPDs depend upon the different longitudinal momentum 
fractions $x + \xi$ ($x - \xi$) of the initial (final) quarks 
(see upper left panel of Fig.~\ref{fig:diagrams}).
As the momentum fractions of the initial and final quarks are different, 
in contrast to the forward parton distributions, 
one accesses in this way quark momentum correlations in the nucleon, 
which are at present largely unknown. 
Furthermore, sum rule integrals of GPDs over $x$ provide new nucleon
structure information and are also amenable to lattice QCD
calculations for direct comparison. 
In particular, the second moment of a particular combination of GPDs 
gives access to the total angular momentum carried by quarks in the
nucleon \cite{Ji97}. Such a quantity would be highly complementary to the
information extracted from polarized deep-inelastic scattering
experiments, which found that about 20 - 30 \% of the nucleon spin 
originates from the quark intrinsic spins (see Ref.~\cite{FJ01} 
for a recent review). 
\newline
\indent
To obtain these new informations,  
one of the main challenges is to directly extract the GPDs from
observables. In the DVCS 
or hard exclusive meson electroproduction observables, the  
GPDs enter in general in convolution integrals over the average 
quark momentum fraction $x$, so that only $\xi$ (half the 
difference of both quark momentum fractions)  
can be accessed experimentally.
A particular exception is when one measures an observable 
proportional to the imaginary part of the amplitude, 
such as the beam helicity asymmetry in DVCS. Then, one 
actually measures directly the GPDs at some specific point, $x=\xi$, 
which is certainly an important gain of information
but clearly not sufficient to map out the GPDs independently in 
both quark momentum fractions, which is needed to construct sum rules.
In absence of any model-independent ``deconvolution" procedure 
at this moment, existing analyses of DVCS experiments 
have to rely on some global model fitting procedure.
\newline
\indent
The double DVCS (DDVCS) process, i.e. the scattering of a spacelike
virtual photon from the nucleon with the production of a 
virtual photon in the final state, provides a way around 
this problem of principle. Compared to the DVCS process with a real
photon in the final state, the virtuality of the 
final photon in DDVCS yields an additional lever arm,
which allows to vary both quark momenta $x$ and $\xi$ independently. 
This additional lever arm of the DDVCS 
compared to the DVCS process has already been noted in 
Refs.~\cite{Mul94,BR00,BMK02}. Also,   
the lepton pair production process induced by a
real photon, i.e. the $\gamma p \to e^+ e^- p$ reaction has been
studied~\cite{BDP02}.
In this letter, we provide the first numerical estimates of the DDVCS and its
competing processes and show how the GPDs can be directly
extracted from DDVCS observables. 
\newline
\indent
The DDVCS process can be accessed through the $l p \to l p e^+ e^-$
reaction, which is characterized by the four-momenta : $k$ ($k'$) of 
the incoming (scattered) leptons $l$, $p$ ($p'$) of the initial (final)
nucleons, and $l_e^-, l_e^+$ of the leptons in the
produced $e^+ e^-$ pair. 
\newline
\indent
To see how the DDVCS process can yield more complete information of
GPDs than the DVCS process, one first has to discuss its richer
kinematics. The DDVCS process is characterized by 8 independent
kinematical variables. Firstly, there are the same 5 kinematical
variables which specify the DVCS process and which we choose as~:
the initial beam energy $E_e$; 
the virtuality $Q^2$ of the incoming photon in the upper left diagram
of Fig.~\ref{fig:diagrams}, i.e. $Q^2 \equiv - q^2$, where $q \equiv k - k'$; 
the usual Bjorken variable $x_B \equiv Q^2 / (2 p . q)$;
the four-momentum transfer to the nucleon $t \equiv \Delta^2$, 
where $\Delta \equiv p' - p$; 
and the out-of-plane angle $\Phi$ between the production plane,
spanned by the vectors $\vec q$ and $\vec q \, '$, and the scattering
plane spanned by the vectors $\vec k$ and $\vec k \, '$. 
Furthermore, one needs 3 additional variables to fully
characterize the DDVCS process which we choose as~:
the virtuality $q'^2 \equiv (l_{e^-} + l_{e^+})^2$ 
of the produced $e^+ e^-$ pair; and the 
2 angles of one lepton of the produced lepton pair, 
evaluated in the {\it c.m.} system of the $e^+ e^-$ pair, and which
span the solid angle $ d \Omega_{e^-}^{*}$. 
\newline
\indent
At large $Q^2$, we calculate the DDVCS process in the handbag
approximation as shown in Fig.~\ref{fig:diagrams} (upper left
diagram), which yields the amplitude~:
\begin{eqnarray}
H^{\mu \nu }_{\mathrm{DDVCS}} &&=\, 
{1\over 2}\, (-g^{\mu \nu})_\perp \, \nonumber \\
\times && \hspace{-.4cm} \int _{-1}^{+1}dx \; C^+(x, \xi, \xi^\prime) 
\left[ \, H^{p}(x,\xi ,t)\, \bar{N}(p^{'}) \, {\Dirac n} \, N(p)\right.
\nonumber\\
&&\left. \hspace{1.2cm} +\, E^{p}(x,\xi ,t)\,
\bar{N}(p^{'})i\sigma ^{\kappa \lambda }
{{n_{\kappa }\Delta _{\lambda }}\over {2m_{N}}}N(p)\right] \nonumber \\
+ &&\, {i\over 2}\, (\epsilon^{\nu \mu})_\perp \nonumber \\
\times && \hspace{-.4cm} \int _{-1}^{+1}dx \; C^-(x,\xi, \xi^\prime) 
\left[\, \tilde{H}^{p}(x,\xi ,t) \,
\bar{N}(p^{'})\, {\Dirac n} \gamma_{5} \, N(p) \right.\nonumber\\
&&\left. \hspace{1.2cm}+\,\tilde{E}^{p}(x,\xi ,t) \,
\bar{N}(p^{'})\gamma_{5}{{\Delta \cdot n}\over {2m_{N}}}N(p) \right], 
\label{eq:ddvcsampl}
\end{eqnarray}
where $\mu$ ($\nu$) refer to the four-vector indices of the incoming
spacelike (outgoing timelike) virtual photons respectively, $n$ is a
light-like vector along the direction of the incoming virtual photon, and  
where we refer to Ref.~\cite{GPV01} for 
the expressions of the symmetrical (antisymmetrical) twist-2 tensors 
$g^{\mu \nu}_\perp$ ($\epsilon^{\mu \nu}_\perp$ ). 
Furthermore, in Eq.~(\ref{eq:ddvcsampl}), $N(p), N(p')$ represent the nucleon
spinors and $m_N$ is the nucleon mass.
The GPDs $H, E, \tilde H, \tilde E$ in Eq.~(\ref{eq:ddvcsampl}) 
are the same as in the DVCS case, and depend on the arguments $x$, $\xi$,
and $t$, with $x$ and $\xi$ as defined in Fig.~\ref{fig:diagrams}.
The coefficient functions $C^\pm$ in the DDVCS amplitude 
of Eq.~(\ref{eq:ddvcsampl}) take the form~:
\begin{eqnarray}
C^\pm(x,\xi,\xi^\prime)=\frac{1}{x- \left(2 \xi^\prime - \xi \right) 
+i\varepsilon}\pm
\frac{1}{x+ \left(2 \xi^\prime - \xi \right)-i\varepsilon} ,
\label{eq:alf}
\end{eqnarray}
where $-2 \xi'$ and $2 (\xi - \xi^\prime)$ are the 
longitudinal momentum fractions  
of the incoming spacelike and outgoing timelike virtual photons
respectively (see Fig.~\ref{fig:diagrams}). 
In the large $Q^2$ limit, one has $2 \xi' \to x_B / (1 - x_B/2)$.
The difference $(2\xi^\prime-\xi)$ appearing in the quark propagators
in Eq.~(\ref{eq:alf}) can be expressed as (relative to $\xi$)~: 
\begin{eqnarray}
{{2 \xi' - \xi} \over {\xi}} \,&=&\, 
{{1 - (q'^2 - \Delta^2)/Q^2 + 8 \xi'^2 \bar m^2 / Q^2} \over 
{1 + (q'^2 - \Delta^2)/Q^2}} \nonumber\\
&\to& {{1 - q'^2/Q^2} \over {1 + q'^2/Q^2}} ,
\label{eq:range}
\end{eqnarray}
with $\bar m^2 = m_N^2 - \Delta^2 / 4$.
For the DDVCS process, by varying the virtualities of 
both incoming and outgoing virtual photons, one can vary independently
the variables, $\xi$ and $\xi^\prime$, whereas, 
in DVCS, only one variable can be varied as $\xi \approx \xi'$. 
One then sees from Eqs.~(\ref{eq:ddvcsampl},\ref{eq:alf})
that the imaginary part of the DDVCS 
amplitude (which can be directly measured through 
the beam helicity asymmetry as discussed further on) 
will access, in a concise notation,
the GPD$(2\xi^\prime-\xi,\xi,t)$, and allows to map out the GPDs as function of
its three arguments independently.
In the second line in Eq.~(\ref{eq:range}), we have indicated the
expression in the large $Q^2$ limit, which is displayed in
Fig.~\ref{fig:range}. 
For a timelike virtual photon (i.e. $q^{\prime 2} > 0$), 
one can only access the $x < \xi$ region in the arguments $(x, \xi, t)$ of
the GPDs because, for kinematical reasons, 
$q'^2 / Q^2$ will always be less than 1. 
Therefore, the imaginary part of the DDVCS amplitude maps out the
GPD where its first argument lies in the range $0 < 2\xi^\prime-\xi <
\xi$. In particular, when $q'^2$ is varied from 0 to $Q^2 / 2$, the
argument $x$ spans about 2/3 of the range $[x, \xi ]$. 
Although one does not access the whole range in $x$, 
clearly, the gain of information on the GPDs
is tremendous as no deconvolution is involved to access this region of
the GPDs. Furthermore, $x < \xi$ is just the range where the GPDs 
contain wholly new information on mesonic ($q \bar q$) components of
the nucleon, which is absent in the forward limit (where $\xi = 0$). 
To access the range $x > \xi$ one would need two spacelike virtual
photons, necessitating to select the two-photon exchange process in
elastic electron nucleon scattering.  
\newline
\indent
Besides the DDVCS process, the $l p \to l p e^+ e^-$ reaction 
contains two classes of Bethe-Heitler (BH) processes as shown in 
Fig.~\ref{fig:diagrams}. 
The BH processes are fully calculable as they involve elastic
nucleon form factors. 
Furthermore, the outgoing timelike photon
which couples pointlike to the quark line in the DDVCS process can
also originate from a neutral vector meson (VM) which couples to the
quark line through a one-gluon exchange (upper right diagram
in Fig.~\ref{fig:diagrams}). For the contamination of the VM
production, we estimate it by the leading order amplitude for the hard
electroproduction of longitudinally polarized VM \cite{VGG98}. 
For this process, which is of order ${\cal O}(\alpha_s)$ 
in the strong coupling constant compared to the handbag process,  
a factorization theorem has been proved \cite{Col97}, allowing to
express its amplitude also in terms of GPDs.
\newline
\indent
In the following, we will estimate the coherent sum of all these
processes. The fully differential cross section of the 
$l p \to l p e^+ e^-$ reaction can then be expressed as~: 
\begin{eqnarray}
&&\hspace{-.4cm} {{d \sigma} \over {d Q^2 d x_B d t \, d \Phi d q'^2 
d \Omega_{e^-}^{*} }} = 
{1 \over {(2 \pi)^4}} \cdot {{x_B  \, y^2} \over {32 \, Q^4
\left( 1 + {{4 m_N^2 x_B^2} \over {Q^2}}\right)^{1/2} }}
\nonumber \\
&& \hspace{2cm} \times {1 \over {(4 \pi)^3}} \; 
\biggl| \, T_{BH} + T_{DDVCS} + T_{VM} \, \biggr|^2 \, ,
\label{eq:crossddvcs}
\end{eqnarray}
where $y \equiv (p . q)/(p . k)$, and where $T_{BH}$, $T_{DDVCS}$, and
$T_{VM}$ are the amplitudes for the BH, DDVCS and VM processes
respectively. 
When integrating Eq.~(\ref{eq:crossddvcs}) 
over the angles of the produced $e^+ e^-$ pair, the
resulting DDVCS cross section reduces in the limit 
$q'^2 \to 0$ to~:
\begin{eqnarray}
{{d \sigma} \over {d Q^2 \, d x_B \, d t \, d \Phi \, d q'^2}} \, \to \,
\left( {{d \sigma} \over {d Q^2 \, d x_B \, d t \, d \Phi}} \right) 
\cdot {N \over {q'^2}} \, ,
\label{eq:crossddvcs2}
\end{eqnarray}
where the DVCS cross section appears on the {\it rhs} 
of Eq.~(\ref{eq:crossddvcs2}). 
The factor $N$ in Eq.~(\ref{eq:crossddvcs2}) is given by 
\begin{eqnarray}
N \,=\, {{\alpha_{em}} \over {4 \pi}} \cdot {4 \over 3} \, , 
\label{eq:N}
\end{eqnarray}
where $\alpha_e\approx 1/137$ is the fine structure constant, 
introduced by the decay of the outgoing photon into the lepton pair.
One sees that the downside of the DDVCS process is that it involves 
small cross sections : at a virtuality $q'^2$ = 1 GeV$^2$, the DDVCS cross
section is reduced by at least a factor $N^{-1}$ ($\approx 1.3 \times 10^3$)
compared to the DVCS cross section. At lower values of $q'^2$, the
DDVCS cross section rises however as $1/q'^2$. 
\newline
\indent
Besides the DDVCS cross section, a particularly informative observable
is obtained by scattering a longitudinally polarized lepton beam and
flipping its helicity. The resulting single spin asymmetry (SSA) 
originates from the interference of the DDVCS and BH processes as~:
\begin{eqnarray}
SSA \sim  {\mathrm{Im}}  \left[ T_{BH} \left( T_{DDVCS} + T_{VM} \right)^*
\right] \, .
\label{eq:ddvcsssa}
\end{eqnarray}
Because the BH process is real, the SSA accesses the imaginary part of
the DDVCS + VM process, which is proportional to the 
GPD($2 \xi^\prime - \xi, \xi, t$) (see Eqs.~(\ref{eq:ddvcsampl},\ref{eq:alf})).
\newline
\indent
\begin{figure}[h]
\vspace{-.4cm}
\includegraphics[width=8.5cm]{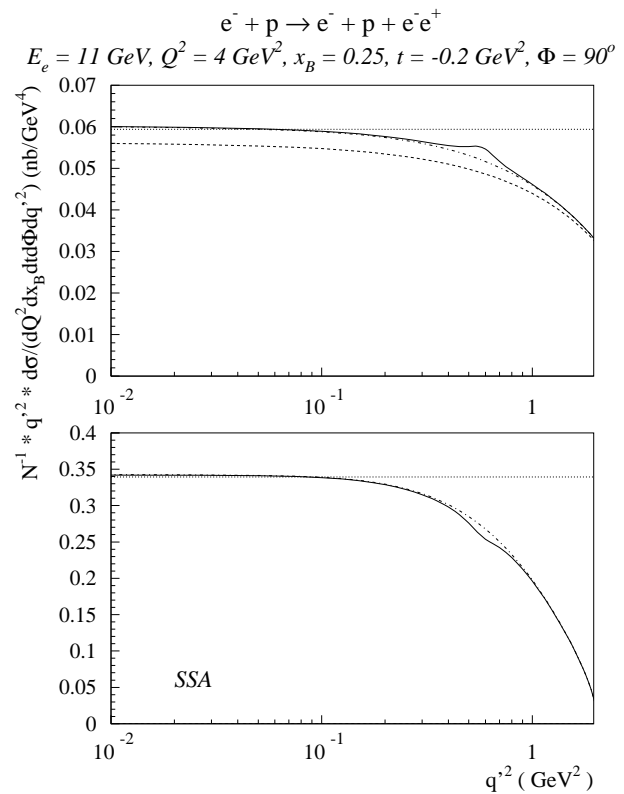}
\vspace{-.4cm}
\caption[]{Cross section (upper panel) and SSA (lower panel) of the 
$e p \to e p e^+ e^-$ process as function of the $e^+ e^-$
virtuality $q'^2$. Dashed curves : BH processes;
dashed-dotted curves : BH + DDVCS processes, full curves : BH + DDVCS
+ $\rho^0_L$ processes. The dotted curves are the corresponding results for
the $e p \to e p \gamma$ proces. The $e p \to e p e^+ e^-$ 
cross section is scaled with $N^{-1} \cdot q'^2$, 
in order to reproduce exactly the $e p \to e p \gamma$ cross section
in the limit $ q'^2 \to 0$, according to Eq.~(\ref{eq:crossddvcs2}).}
\label{fig:ssa11_ddvcs}
\end{figure}
In Fig.~\ref{fig:ssa11_ddvcs}, we show the $q'^2$ dependence of the 
estimated cross section and SSA for the $e p \to e p e^+ e^-$ process 
in kinematics accessible at JLab.
As the twist-2 SSA basically displays a $\sin \Phi$ structure, we show
its value at $\Phi = 90^o$.
For the GPDs, we use a $\xi$-dependent parametrization 
(see Refs.~\cite{VGG98,GPV01}), using the MRST01 \cite{MRST01}
forward quark distributions as input. 
As is seen from Fig.~\ref{fig:ssa11_ddvcs}, we firstly confirm 
numerically that
the $e p \to e p e^+ e^-$ cross section scaled with the factor 
$N^{-1} q'^2$ reduces to the $e p \to e p \gamma$ cross section 
when approaching the real photon point. Similarly, the SSA for the 
$e p \to e p e^+ e^-$ process reduces to the corresponding SSA for the 
$e p \to e p \gamma$ process. When going to larger virtualities
$q'^2$, the DDVCS shows a growing deviation from the $1/q'^2$ behavior
and the magnitude of the SSA decreases. Furthermore, we show in 
 Fig.~\ref{fig:ssa11_ddvcs} the contribution of the $\rho^0_L \to e^+
e^-$ process (upper right diagram of Fig.~\ref{fig:diagrams}), 
which is the most pronounced VM process. We find that, except in the 
immediate vicinity of $q'^2 \approx m_\rho^2$, the $\rho^0_L \to e^+
e^-$ process is very small. This can be understood because the cross
section for the $\rho^0_L \to e^+ e^-$ process is reduced by a factor 
$\alpha_{em}^2$ compared to the $\rho^0_L$ process, 
whereas the DDVCS cross section is only reduced by a factor 
$\alpha_{em}$ compared to the DVCS cross section.
Similarly, the SSA is only slightly affected by the VM process
and is dominantly proportional to the imaginary part of the DDVCS
process according to Eq.~(\ref{eq:ddvcsssa}). The strong sensitivity
of the SSA on $q'^2$, as seen from Fig.~\ref{fig:ssa11_ddvcs}, should
therefore allow to map out the GPDs in the range $x < \xi$.
\newline
\indent
Fig.~\ref{fig:ddvcs6_qps} shows the 
comparison between the $e p \to e p (\gamma, \rho^0_L)$ and  
$e p \to e p (\gamma, \rho^0_L) \to e p (e^+ e^-)$ 
processes for a typical kinematics accessible at JLab at 6 GeV. 
Whereas the $e p \to e p \rho^0_L$ process is roughly 
comparable to the DVCS + BH one for these kinematics,
their ``timelike'' analogues, show that the $\rho^0_L$ channel 
is suppressed by 2 orders of magnitude with respect to the DDVCS+BH.
\newline
\indent
Given that about $10^4$ DVCS+BH events were 
measured recently at CLAS \cite{Step01} in an effective 
4-day data taking period at a luminosity of $10^{34} cm^{-2}s^{-1}$ in
a non-dedicated experiment, it can certainly be envisaged that 
the DDVCS+BH cross section, which is about 3 orders 
of magnitude lower, be measured with a dedicated long-time experiment. 
A luminosity of $10^{35} cm^{-2} s^{-1}$, projected at CLAS for the upgrade of 
JLab at 12 GeV, or at a future dedicated lepton facility, would 
allow to measure this reaction with reasonable statistics. 
\newline
\indent
In conclusion, we have studied the DDVCS process with the production
of a timelike virtual photon, decaying into an $e^+ e^-$ pair. We have
expressed the DDVCS amplitude in terms of GPDs and have shown that 
by varying the virtuality $q'^2$ of the timelike photon, one can map
out the GPDs as function of both initial and final quark momentum
fractions.  We have given cross section estimates
for the DDVCS and its associated processes. 
Although the cross sections are small, their measurement seems feasible 
with a dedicated experiment at JLab and at a future
high-energy, high-luminosity lepton facility.
Of particular interest is the SSA using a polarized lepton beam. 
We have shown that by measuring the SSA, one can 
directly extract the GPDs in the domain where one is sensitive 
to $q \bar q$ correlations in the nucleon, providing a whole new source of 
nucleon structure information which is absent in forward quark distributions.

This work was supported by the Deutsche Forschungsgemeinschaft (SFB443), 
by the CNRS/IN2P3, 
and in part by the European Commission IHP program 
(contract HPRN-CT-2000-00130).

\end{document}